\documentclass[12pt]{article}%
\usepackage{amssymb}
\usepackage{amsmath}
\usepackage{epsfig}
\usepackage{amsfonts}
\usepackage{graphicx}
\usepackage{hyperref}%
\providecommand{\U}[1]{\protect\rule{.1in}{.1in}}
\begin{document}

\title{Scattering Amplitudes for Multi-indexed Extensions of Soliton Potential and
Extended KdV Integer Solitons}
\author{Jen-Chi Lee\\Department of Electrophysics,National Chiao-Tung University and\\Physics Division, National Center for Theoretical Sciences,\\Hsinchu, Taiwan, R.O.C.\\{jcclee@cc.nctu.edu.tw}}
\date{\today }
\maketitle

\begin{abstract}
We calculate quantum mechanical scattering problems for multi-indexed
extensions of soliton potential by Darboux transformations in terms of pseudo
virtual wavefunctions. As an application, we calculate infinite set of higher
integer KdV solitons by the inverse scattering transform method of KdV equation.

\end{abstract}

%

\setcounter{equation}{0}
\renewcommand{\theequation}{\arabic{section}.\arabic{equation}}%

\section{\bigskip Introduction}

The discovery of the exceptional orthogonal polynomials
\cite{Canada,Canada2,Que,Que2,OS0,OS-0,OS1}, and their applications to exactly
solvable 1D quantum mechanical systems \cite{Que,Que2,OS1} have recently
attracted many substantial researches in mathematical physics. In this talk,
instead of the bound state problems, we will report on new development of 1D
exactly solvable quantum mechanical scattering problems. In particular, we
will calculate the scattering problems of the deformed soliton potential and
apply the results to the KdV solitons.

Our calculation will be based on the multiple Darboux-Crum transformations
\cite{Dar,Crum,Russ}. Such transformations can generate new solvable quantum
systems from the known ones by using certain polynomial type seed solutions.
These seed functions are called the virtual and pseudo virtual state
wavefunctions \cite{OS1,OS2,OS3}. One way to obtain these seed solutions is to
perform discrete symmetry operations on the eigenfunctions. The one-indexed
\cite{YKM} and more complete multi-indexed extensions \cite{hls} of the known
quantum scattering problems \cite{KS} were recently calculated.

For the case of the deformed soliton potentials, one interesting application
is to use the results of the scattering data to generate higher integer KdV
solitons from the lower integer ones. Although the KdV solitons generated by
this method are not new soliton solutions, we believe that the method we used
is simple and effective, and is closely related to the recent developed
multi-indexed extention of solvable quantum mechanical potentials. In this
calculation, we will obtain an infinite number of reflectionless potentials,
which can be served as the initial profiles of integer KdV solitons. We then
use the scattering data to solve the Gel'fand-Levitan-Marchenko (GLM) equation
\cite{GL,M} in the inverse scattering transform (IST) method \cite{GGKM,DJ},
and obtain higher integer KdV solitons.

\section{The Deformed Soliton Potential}

We begin with a specific example of new solvable potential, namely, the
deformed soliton potential under Darboux-Crum transformation. We will
calculate both its bound state problem and scattering problem, and relate
these two calculations. The bound state problem and scattering problem of the
original soliton potential%

\begin{equation}
U(x;\mathbf{\lambda})=-\frac{h(h+1)}{\cosh^{2}x},\mathbf{\lambda
}=h,h>0,-\infty<x<\infty
\end{equation}
were calculated in \cite{Landau}. This potential contains finitely many bound
states%
\begin{align}
\phi_{n}(x)  &  =\frac{1}{(\cosh x)^{h-n}}P_{n}^{(h-n,h-n)}(\tanh
x)\nonumber\\
&  \sim\frac{1}{(\cosh x)^{h-n}}\text{ }_{2}F_{1}\left(  -n,2h-n+1,h-n+1,\frac
{1-\tanh x}{2}\right)  ,\nonumber\\
E_{n}  &  =-k^{2}=-(h-n)^{2};n=0,1,2...[h]^{\prime}%
\end{align}
where $P_{n}^{(h-n,h-n)}$ is the Jacobi polynomial and $_{2}F_{1}$ is the
hypergeometric function. $[h]^{\prime}$ denotes the greatest integer not
exceeding and not equal to $h$. The soliton potential contains a discrete
symmetry%
\begin{equation}
h\rightarrow-(h+1),
\end{equation}
which can be used to construct the seed function%
\begin{equation}
\varphi_{v}(x)=(\cosh x)^{h+1+v}P_{v}^{(-h-1-v,-h-1-v)}(\tanh
x),v=0,1,2,3,4,..
\end{equation}
with energy $E_{v}=-(h+v+1)^{2}$. It turns out that for $v=1,3,5....$, the
deformed potential contains pole at $x=0$. For example, for $v=1,$%
\begin{equation}
U_{1}=U-2\frac{d^{2}}{dx^{2}}\log\varphi_{1}(x)=U-\frac{2(h+1)}{\cosh^{2}%
x}+\frac{2}{\sinh^{2}x}%
\end{equation}
which contains pole at $x=0$. We note that although one can define the
asymptotic forms of the scattering state for this potential, the corresponding
bound state wavefunctions contain singularities. So for our purpose here, only
$v=2,4,6...$can be used to deform the soliton potential. For simplicity, we
will use the seed function for $v=2$%
\begin{equation}
\varphi_{2}(x)=\frac{h+1}{4}(\cosh x)^{h+3}[1+(2h+3)\tanh^{2}x]
\end{equation}
to illustrate the calculation. The deformed potential is%
\begin{align}
U_{2}  &  =U-2\frac{d^{2}}{dx^{2}}\log\varphi_{2}(x)=U-\frac{4(2h+3)(1-2\sinh
^{2}x)}{[1+(2h+3)\tanh^{2}x](\cosh x)^{4}}\nonumber\\
&  -8\left(  \frac{(2h+3)\tanh x}{[1+(2h+3)\tanh^{2}x](\cosh x)^{2}}\right)
^{2}, \label{deform}%
\end{align}
which has no pole for the whole regime of $x$ and approaches $0$
asymptotically for $x\rightarrow\pm\infty$ as $U$ does. Note that
$U_{2}(x=0)-U(x=0)=-4(2h+3)<0,$ which suggests the existence of a lowest new
bound state for the deformed potential $U_{2}.$

\bigskip The bound state wavefunctions of the deformed potential
Eq.(\ref{deform}) can be calculated through the Darboux-Crum transformation to
be%
\begin{align}
\psi_{b}^{(1)}(x)  &  =\phi_{n}^{\prime}-\frac{\varphi_{2}^{\prime}}%
{\varphi_{2}}\phi_{n}\nonumber\\
&  =-\left(  \frac{(2h-n+3)\tanh x}{(\cosh x)^{h-n}}+\frac{2(2h+3)\tanh
x}{[1+(2h+3)\tanh^{2}x](\cosh x)^{h-n-2}}\right) \nonumber\\
&  \cdot\text{ }_{2}F_{1}\left(  -n,2h-n+1,h-n+1,\frac{1-\tanh x}{2}\right)
\nonumber\\
&  +\frac{n(2h-n+1)}{2(h-n-1)(\cosh x)^{h-n-2}}\nonumber\\
&  \cdot\text{ }_{2}F_{1}\left(  -n+1,2h-n+2,h-n+2,\frac{1-\tanh x}{2}\right)
\label{bound}%
\end{align}
with energy%
\begin{equation}
E_{n}=-k^{2}=-(h-n)^{2};n=0,1,2...[h]^{\prime}.
\end{equation}
In calculating Eq.(\ref{bound}), we have used the identity%
\begin{equation}
\frac{d}{dz}\text{ }_{2}F_{1}\left(  a,b,c,z\right)  =\text{ }_{2}F_{1}\left(
a+1,b+1,c+1,z\right)  .
\end{equation}
It can be easily shown that there is another bound state%
\begin{equation}
\frac{1}{\varphi_{2}}=\frac{4/(h+1)}{(\cosh x)^{h+3}[1+(2h+3)\tanh^{2}x]}%
\end{equation}
with lowest energy%
\begin{equation}
E=-(h+3)^{2} \label{pole}%
\end{equation}
as was expected previously. \qquad\qquad

We next consider the scattering problem of the deformed potential. For this
purpose, we introduce the wave vector $K$ such that%
\begin{equation}
E=K^{2}>0,K=ik.
\end{equation}
In view of the bound state wavefunction in Eq.(\ref{bound}), the scattering
state wavefunction for the undeformed potential is%
\begin{equation}
\psi_{+}(x)=\frac{1}{(\cosh x)^{-iK}}\text{ }_{2}F_{1}\left(
-iK-h,-iK+h+1,-iK+1,\frac{1-\tanh x}{2}\right)  . \label{scatt}%
\end{equation}
The asymptotic form of the scattering state for the deformed potential can be
calculated to be%
\begin{equation}
\psi_{+\infty}^{(1)}(x)=\psi_{+\infty}^{\prime}(x)-\frac{\varphi_{2}^{\prime}%
}{\varphi_{2}}\psi_{+\infty}^{\prime}(x)=[iK-(h+3)]\exp iKx
\end{equation}
as $x\rightarrow+\infty.$ To calculate the asymptotic form of the scattering
state as $x\rightarrow-\infty$, one needs to include the second solution of
the Schrodinger equation. One way to achieve this is to use the identity
\cite{Landau}%
\begin{align}
\text{ }_{2}F_{1}\left(  a,b,c,z\right)   &  =\frac{\Gamma(c)\Gamma
(c-a-b)}{\Gamma(c-a)\Gamma(c-b)}\text{ }_{2}F_{1}\left(
a,b,a+b-c+1,1-z\right) \nonumber\\
&  +\frac{\Gamma(c)\Gamma(a+b-c)}{\Gamma(a)\Gamma(b)}(1-z)^{c-a-b}\text{
}\nonumber\\
&  \cdot_{2}F_{1}\left(  c-a,c-b,c-a-b+1,1-z\right)
\end{align}
in Eq.(\ref{scatt}) to obtain%
\begin{align}
\psi_{-}(x)  &  =\frac{1}{(\cosh x)^{-iK}}\frac{\Gamma(c)\Gamma(c-a-b)}%
{\Gamma(c-a)\Gamma(c-b)}\text{ }_{2}F_{1}\left(  a,b,a+b-c+1,1-z\right)
\nonumber\\
&  +\frac{1}{(\cosh x)^{-iK}}\frac{\Gamma(c)\Gamma(a+b-c)}{\Gamma(a)\Gamma
(b)}(1-z)^{c-a-b}\text{ }\nonumber\\
&  \cdot_{2}F_{1}\left(  c-a,c-b,c-a-b+1,1-z\right)  \label{asymp}%
\end{align}
where $a=-iK-h,b=-iK+h+1,c=-iK+1$ and $z=\frac{1-\tanh x}{2}$. One can now
calculate the asymptotic form of Eq.(\ref{asymp})%
\begin{equation}
\psi_{-\infty}(x)=\exp-iKx\text{ }\frac{\Gamma(-iK+1)\Gamma(iK)}%
{\Gamma(1+h)\Gamma(-h)}\text{ }+\exp iKx\text{ }\frac{\Gamma(-iK+1)\Gamma
(-iK)}{\Gamma(-iK-h)\Gamma(-iK+h+1)}\text{ }%
\end{equation}
as $x\rightarrow-\infty.$ The asymptotic form of the scattering state for the
deformed potential can then be calculated to be%
\begin{align}
\psi_{-\infty}^{(1)}(x)  &  =\psi_{-\infty}^{\prime}(x)-\frac{\varphi
_{2}^{\prime}}{\varphi_{2}}\psi_{-\infty}^{\prime}(x)=[-iK+(h+3)]\exp
-iKx\text{ }\frac{\Gamma(-iK+1)\Gamma(iK)}{\Gamma(1+h)\Gamma(-h)}\text{
}\nonumber\\
&  +[iK+(h+3)]\exp iKx\text{ }\frac{\Gamma(-iK+1)\Gamma(-iK)}{\Gamma
(-iK-h)\Gamma(-iK+h+1)}%
\end{align}
as $x\rightarrow-\infty.$ Finally the transmission and reflection coefficients
of the deformed soliton potential can be calculated to be%
\begin{align}
t_{D}(K)  &  =\frac{iK-(h+3)}{iK+(h+3)}\cdot\frac{\Gamma(-iK-h)\Gamma
(-iK+h+1)}{\Gamma(-iK+1)\Gamma(-iK)}=\frac{iK-(h+3)}{iK+(h+3)}\cdot
t(K),\label{amp}\\
r_{D}(K)  &  =(-)\frac{K+i(h+n+1)}{K-i(h+n+1)}\cdot\frac{\Gamma(1+h-iK)\Gamma
(-h-iK)\Gamma(ik)}{\Gamma(-h)\Gamma(1+h)\Gamma(-iK)}\nonumber\\
&  =(-)\frac{K+i(h+n+1)}{K-i(h+n+1)}\cdot r(K) \label{ref}%
\end{align}
where $t(K)$ and $r(K)$ are the transmission and reflection coefficients of
the undeformed soliton potential. It is important to note that, in
Eq.(\ref{amp}), in addition to the well known second factor for the undeformed
soliton potential scattering, an additional ratio $\frac{iK-(h+3)}{iK+(h+3)}$
with a new pole $K=i(h+3)$ corresponding to the new bound state energy given
by Eq.(\ref{pole}) is added to the transmission coefficient of the deformed
potential. Moreover, the additional ratio is of module one so that the
conservation of flux is preserved for the deformed potential. In general, It
turns out that the $M$-step deformed transmission amplitude $t_{D}(K)$ and the
reflection amplitude $r_{D}(K)$ are \cite{hls}
\begin{equation}
t_{\mathcal{D}}(K)=\prod_{j=1}^{M}\frac{K+i\Delta_{d_{j}}^{+}}{K+i\Delta
_{d_{j}}^{-}}\cdot t(K),\quad r_{\mathcal{D}}(K)=(-1)^{M}\prod_{j=1}^{M}%
\frac{K-i\Delta_{d_{j}}^{-}}{K+i\Delta_{d_{j}}^{-}}\cdot r(K) \label{refA}%
\end{equation}
where $\Delta_{d_{j}}^{+}$ and $\Delta_{d_{j}}^{-}$ are the asymptotic
exponents of seed solutions $\varphi_{d_{j}}$%
\begin{equation}
\varphi_{d_{j}}\approx\left\{
\begin{array}
[c]{cl}%
e^{x\Delta_{d_{j}}^{+}} & \qquad x\rightarrow+\infty\\
e^{x\Delta_{d_{j}}^{-}} & \qquad x\rightarrow-\infty
\end{array}
\right.  .
\end{equation}
In view of the multiplicative form of $r_{D}(k)$ in Eq.(\ref{ref}), it is
important to note that, for integer $h=1,2,3....$, the scattering of the
deformed potential remains reflectionless as the undeformed potential due to
the factor $\Gamma(-h)$ in the denominator of $r(k).$ This key observation is
important for generating higher integer soliton solutions in the calculation
of the inverse transform method to be discussed in the next section.

\bigskip

\section{Solvable Higher Integer KdV Solitons}

The nontopological KdV solitons are described by the Korteweg-de Vries (KdV)
equation, i.e.,
\begin{equation}
u_{t}-6uu_{x}+u_{xxx}=0
\end{equation}
in one space $x\in(-\infty,\infty)$ and one time $t>0$ dimension. One of the
important methods to solve nonlinear partial differential equation such as the
KdV equation is the method of the inverse scattering transform(IST)
\cite{GGKM,DJ} invented in 1960's. According to the method of IST, the
solution of the KdV equation is converted to the solution of two simpler
linear equations, namely, the quantum mechanical Schrodinger equation and the
Gel'fand-Levitan-Marchenko (GLM) equation \cite{GL,M}. For KdV soliton
solutions the related Schr\"{o}dinger equation is connected with
reflectionless potentials. For such reflectionless potentials, the reflection
amplitudes vanish, and the corresponding GLM equation is easy to solve. For
the general $N$-soliton solution, one gets $2N$ continuous parameters, $N$
norming constants $c_{n}(0)$ and $N$ energy parameters $\kappa_{n}$.

\bigskip In the following \cite{HL}, for simplicity and clarity of
presentation, we present the result for $1$-step deformation and take the seed
function with $v=2$, and parameter in the soliton potential $h=1$%
\begin{equation}
\varphi_{2}(x)_{h=1}=\frac{1}{2}\cosh^{4}x(1+5\tanh^{2}x)
\end{equation}
to illustrate the calculation. The deformed potential is easily calculated to
be%
\begin{equation}
U_{2}(x)_{h=1}=U(x)-2\frac{d^{2}}{dx^{2}}\log\varphi_{2}(x)_{h=1}%
=-\frac{30(4\cosh^{4}x-8\cosh^{2}x+5)}{\cosh^{2}x(36\cosh^{4}x-60\cosh
^{2}x+25)} \label{U}%
\end{equation}
which has no pole and no zero for the whole regime of $x$ and approaches $0$
asymptotically for $x\rightarrow\pm\infty$ as $U(x)_{h=1}$ does. The bound
state wavefunctions of the deformed potential Eq.(\ref{U}) can be calculated
through the Darboux-Crum transformation to be%
\begin{equation}
\psi_{0}(x)=\phi_{0}^{\prime}-\frac{\varphi_{2}^{\prime}}{\varphi_{2}}\phi
_{0}=-5\,\mathrm{sech}x\tanh x\left(  1+\frac{2\,\mathrm{sech}^{2}x}%
{(1+5\tanh^{2}x)}\right)
\end{equation}
with energy%
\begin{equation}
E_{0}=-\kappa_{0}^{2}=-(h-n)^{2}=-1.
\end{equation}
It can be easily shown that there is another bound state of the deformed
potential%
\begin{equation}
\psi_{1}(x)\sim\frac{1}{\varphi_{2}}=\frac{2}{\cosh^{4}x(1+5\tanh^{2}x)}%
\end{equation}
with a lower energy%
\begin{equation}
E_{1}=-\kappa_{1}^{2}=-(h+1+v)^{2}=-4^{2}%
\end{equation}
as was expected previously. The normalized wavefunctions and their asymptotic
forms can be calculated to be%
\begin{equation}
\psi_{0}(x)=\sqrt{\frac{15}{2}}\,\mathrm{sech}x\tanh x\left(  1+\frac
{2\,\mathrm{sech}^{2}x}{(1+5\tanh^{2}x)}\right)  \rightarrow\sqrt{\frac{10}%
{3}}e^{-x}\text{ as }x\rightarrow\infty, \label{C}%
\end{equation}%
\begin{equation}
\psi_{1}(x)=\sqrt{\frac{15}{8}}\frac{2}{\cosh^{4}x(1+5\tanh^{2}x)}%
\rightarrow\sqrt{\frac{40}{3}}e^{-x}\text{ as }x\rightarrow\infty. \label{CC}%
\end{equation}
The constants%
\begin{equation}
c_{0}(0)=\sqrt{\frac{10}{3}},~~~c_{1}(0)=\sqrt{\frac{40}{3}}%
\end{equation}
in equations Eq.(\ref{C}) and Eq.(\ref{CC}) are called norming constants. The
reflection amplitude of the scattering of the $M$-step ($M=1$ for the present
case) deformed soliton potential Eq.(\ref{U}) was calculated in Eq.(\ref{refA}%
).\ In view of the multiplicative form of $r_{D}(k)$, it is important to note
that, for integer $h=1,2,3....$, the scattering of the deformed potential
remains reflectionless as the undeformed potential due to the factor
$\Gamma(-h)$ in the denominator of $r(k).$

We can now use the scattering data $\{\kappa_{n},c_{n},r_{D}(k)\}$ to solve
the KdV equation. For the reflectionless potential, $r_{D}(k)=0$, the GLM
equation is easy to solve, and the solution $u(x,t)$ is given by \cite{DJ}
\begin{equation}
u(x,t)=-2\frac{d^{2}}{dx^{2}}\log(\det A),
\end{equation}
where $A$ is a $N\times N$ matrix ($N\equiv h+1$) with elements $A_{mn}$ given
by
\begin{equation}
A_{mn}=\delta_{mn}+c_{n}^{2}(t)\frac{\exp-(\kappa_{m}+\kappa_{n})x}{\kappa
_{m}+\kappa_{n}};~~~m,n=0,1,2....,N-1. \label{AA}%
\end{equation}
In Eq.(\ref{AA}) $c_{n}(t)=c_{n}(0)\exp(4\kappa_{n}^{3}t)$ and is one of the
Gardner-Greene-Kruskal-Miura (GGKM) equations \cite{GGKM}.

For the present case, $N=h+1=2$. The integer $2$-soliton solution
corresponding to $(\kappa_{0,}\kappa_{1})=(1,4)$ can be calculated to be
\cite{HL}%
\begin{equation}
u(x,t)_{(1,4)}=-\frac{120e^{8t+2x}(e^{1024t}+e^{16x}+16e^{520t+6x}%
+30e^{512t+8x}+16e^{504t+10x})}{(3e^{520t}+3e^{10x}+5e^{512t+2x}%
+5e^{8t+8x})^{2}}. \label{1,4t}%
\end{equation}
By taking $t=0$ in Eq.(\ref{1,4t}), one reproduces the initial profile
$u(x,0)=U_{2}(x)_{h=1}$ calculated in Eq.(\ref{U}). The asymptotic form of the
$(\kappa_{0,}\kappa_{1})=(1,4)$ solution is
\begin{equation}
u(x,t)_{(1,4)}\sim-2\sum_{n=0}^{N-1}\kappa_{n}^{2}\sec h^{2}\{\kappa
_{n}(x-4\kappa_{n}^{2}t)\pm\chi_{n}\},t\rightarrow\pm\infty,
\end{equation}
where%
\begin{equation}
\exp(2\chi_{n})=\prod_{%
\genfrac{}{}{0pt}{}{m=0}{{m\neq n}}%
}^{N-1}\left\vert \frac{\kappa_{n}-\kappa_{m}}{\kappa_{n}+\kappa_{m}%
}\right\vert ^{sgn(\kappa_{n}-\kappa_{m})}.
\end{equation}
Note that the previous integer 2-soliton solution corresponds to $(\kappa
_{0,}\kappa_{1})=(1,2)$. It is interesting to see that the calculation of the
$(1,4)$ integer soliton is based on the recently developed multi-indexed
extensions of the reflectionless soliton potential.

\bigskip The calculation can be generalized to other higher integer solitons.
The profile for $h=2$, for example, is the extended solvable $3$-soliton
$(1,2,5)$ \cite{HL}%
\begin{equation}
U_{2}(x)_{h=2}=u(x,0)_{(1,2,5)}=-\frac{4(144\cosh^{4}x-280\cosh^{2}%
x+147)}{\cosh^{2}x(64\cosh^{4}x-112\cosh^{2}x+49)},
\end{equation}
and
\begin{align}
u(x,t)_{(1,2,5)}  &  =-(16e^{8t+2x}(9e^{2128t}+9e^{28x}+1575e^{16(63t+x)}%
+882e^{16(66t+x)}\nonumber\\
&  +3252e^{14(76t+x)}+175e^{8(142t+x)}+49e^{8(250t+x)}+126e^{4(516t+x)}%
\nonumber\\
&  +56e^{2072t+2x}+126e^{2056t+6x}+1008e^{1128t+10x}+882e^{1072t+12x}%
\nonumber\\
&  +1575e^{1120t+12x}+1008e^{1000t+18x}+49e^{128t+20x}\nonumber\\
&  +175e^{992t+20x}+126e^{72t+22x}+126e^{64t+24x}+56e^{56t+26x})\nonumber\\
&  /(2e^{1072t}+2e^{16x}+14e^{4(252t+x)}+9e^{2(532t+x)}+7e^{1000t+6x}%
\nonumber\\
&  +7e^{72t+10x}+14e^{64t+12x}+9e^{8t+14x})^{2}.
\end{align}

\section{Discussion}

In this talk we report on the recent calculation of the scattering problems of
multi-indexed extensions of the soliton potential. Similar calculation can be
performed to other known solvable potentials \cite{hls}. As an application
\cite{HL}, we point out an infinite set of higher integer initial profiles of
the KdV solitons, which are both exactly solvable for the Schrodinger equation
and for the Gel'fand-Levitan-Marchenko equation in the inverse scattering
transform method of KdV equation. The calculation of these solitons are based
on the multi-indexed extensions of the reflectionless soliton potential in
terms of the Darboux-Crum transformations.

\section{\bigskip Acknowledgments}

This talk is based on a work with Choon-Lin Ho and Ryu Sasaki \cite{hls}, and
another work with Choon-Lin Ho \cite{HL}. The work is supported in part by

NSC-100-2112-M-009-002-MY3, and S.T. Yau center of NCTU, Taiwan.

\end{document}